\documentclass[aps,prb,twocolumn,amsmath,amssymb,nofootinbib,superscriptaddress,floatfix,eqsecnum]{revtex4}

\usepackage{amsmath}
\usepackage{amssymb}
\usepackage{amsthm}
\usepackage{color}
\usepackage[pdftex]{hyperref} 
\usepackage{graphicx}
\usepackage{dcolumn} 
\usepackage{bm} 
\usepackage{epic}
\usepackage{longtable}
\usepackage{ulem}   
\usepackage{slashed}
\newcommand{\bs}[1]{{\boldsymbol{#1}}}

\begin{document}
\title{Topological BF theory of the quantum hydrodynamics of incompressible
polar fluids}

\author
{
Apoorv Tiwari
}
\affiliation
{
Department of Physics, University of Illinois at Urbana-Champaign, 1110 West Green St, Urbana IL 61801
}
\author
{
Xiao Chen
}
\affiliation
{
Department of Physics, University of Illinois at Urbana-Champaign, 1110 West Green St, Urbana IL 61801
}
\author
{
Titus Neupert
}
\affiliation
{
Princeton Center for Theoretical Science, Princeton University, Princeton, New Jersey 08544, USA
}
\author
{
Luiz H. Santos
}
\affiliation
{
Perimeter Institute for Theoretical Physics, Waterloo, ON, N2L 2Y5, Canada
}
\author
{
Shinsei Ryu
}
\affiliation
{
Department of Physics, University of Illinois at Urbana-Champaign, 1110 West Green St, Urbana IL 61801
}
\author
{
Claudio Chamon
}
\affiliation
{
Physics Department, Boston University, Boston, Massachusetts 02215, USA
}
\author
{
Christopher Mudry
}
\affiliation
{
Condensed Matter Theory Group, Paul Scherrer Institute, CH-5232 Villigen PSI, Switzerland
}

\date{\today}

\begin{abstract}
We analyze a hydrodynamical model of a polar fluid in
(3+1)-dimensional spacetime. We explore a
spacetime symmetry -- volume preserving diffeomorphisms -- to
construct an effective description of this fluid in terms of a
topological BF theory. The two degrees of freedom of the BF theory are
associated to the mass (charge) flows of the fluid and its
polarization vorticities. We discuss the quantization of this
hydrodynamic theory, which generically allows for fractionalized
excitations. We propose an extension of the 
Girvin-MacDonald-Platzman algebra to (3+1)-dimensional spacetime 
by the inclusion of the vortex-density operator 
in addition to the usual charge density operator
and show that the same algebra is obeyed by massive Dirac fermions 
that represent the bulk of $\mathbb{Z}^{\,}_{2}$ topological
insulators in three-dimensional space.
\end{abstract}

\maketitle

\section{Introduction}

One of the most prominent topological phenomena in quantum matter is
the quantum Hall effect.\cite{QHE-Book} It comes about when a two
dimensional electron gas is subject to a strong perpendicular magnetic
field at sufficiently low temperatures. The striking property of this
many-body state, its quantized transport, derives from an
incompressible (i.e., gapped) state in the bulk accompanied by soft
chiral edge modes along the one dimensional boundary. The electrons
from the bulk state can be thought of as giving rise to an incompressible
fluid state.

The incompressible fluid picture of the quantum Hall effect has been
investigated by Bahcall and Susskind,%
~\cite{Bahcall91,Susskind01} 
who have shown that some properties of the quantum Hall state can be
accounted for if one considers a classical two dimensional
incompressible fluid model of charged point particles in a perpendicular
magnetic field and if one applies a semi-classical analysis thereof.

In the construction presented in
Refs.~\onlinecite{Bahcall91}
and \onlinecite{Susskind01}, the fluid description arises
by considering the limit when the inter-particle distance
is sufficiently small. In this limit, 
the individual positions of particles can
be effectively described by a collective coordinate of the fluid. The
freedom to relabel the discrete particles emerges as a gauge symmetry
in the fluid formulation (see, for instance, the review in
Ref.~\onlinecite{Jackiw04}).  
The classical Lagrangian of the fluid of
charged particles contains an (Abelian) Chern-Simons term whose vector
field undergoes a gauge transformation that is equivalent to a
reparametrization of the fluid's underlying particles. Given that the
Chern-Simons action captures the topological essence of the quantum
Hall state, the fluid formulation of the Hall effect discussed in
Refs.~\onlinecite{Bahcall91} and 
\onlinecite{Susskind01} 
offers an insightful platform
for understanding the interplay of incompressibility and topology
as it relates to two dimensional systems in an applied magnetic field.

In recent years, a number of new topological phenomena has arisen that
go beyond the quantum Hall paradigm. In particular, topological band
insulators in two and three dimensions have been predicted and
experimentally found in solid state systems.%
~\cite{Hasan2010,Qi2011}
The discovery of this new class of materials has 
revitalized the interest for non-interacting%
~\cite{Schnyder2008,Kitaev2009,Ryu2010} 
and interacting%
~\cite{Levin2009,Neupert2011,Santos2011,Chen2012,Lu2012,Vishwanath2013,Wang2014}
topological phases of matter.

Motivated by the construction of
Refs.~\onlinecite{Bahcall91}
and \onlinecite{Susskind01}, we propose a fluid model in
three dimensional space whose effective action contains a BF topological
term,~\cite{Blau1991} 
the natural generalization of the Chern-Simons term to three dimensions. 
The new feature of our model, 
aside from the dimensionality three of space, 
is that, in order to obtain a topological BF term, 
we are led to consider a polar incompressible fluid, 
while the fluid is made of point particles in 
Refs.~\onlinecite{Bahcall91}
and \onlinecite{Susskind01}.
We propose a Lagrangian written in the explicit
coordinates of the fluid's particles, 
position and dipole field, 
and show that, by expressing this term as a function of the small
fluctuations of the particle's positions, it renders a topological
BF action.

The BF term captures the Berry phase associated to a point defect
adiabatically winding around a vortex
line.~\cite{Blau1991,Bergeron1995,Szabo1998,Hansson2004,Cho2011}
The Berry phase associated to this adiabatic motion yields 
the statistics between point and vortex defects in three dimensions.
In our formulation, the coefficient of the BF action emerges 
as a function of the phenomenological parameters of the fluid, 
which is subsequently shown to satisfy a quantization condition 
upon quantization of the fluid.

We also find that the BF theory furnishes a pair of conserved currents,
i.e., a charge current and a vorticity current. We interpret these currents
within a massive Dirac model as the usual fermion current and the
fermion ``vorticity'' current respectively. Upon evaluation the
algebra for the projected charge and spin densities in the Dirac model, 
we find that it agrees with the BF algebra. We find an
algebra very similar to the celebrated 
Girvin-MacDonald-Platzman (GMP) algebra for FQH systems,
with the inclusion of a vorticity sector in addition to the charge
sector.

This paper is organized as follows.  In Sec.%
~\ref{sec:Review of Lagrangian fluids}, 
we provide a short review of Lagrangian fluids
focusing on the main aspect related to our work, namely the role
played by the invariance under volume preserving diffeomorphisms.  In
Sec.~\ref{sec:BF Lagrangian for an incompressible polar fluid}, we
propose a classical model for a polar incompressible fluid, which
leads to BF term effective action once small fluctuations of the fluid
are take into account. Quantization of this fluid leads to a the
identification of the quasi-particles (point-like and vortex-like) as
well as their mutual statistics determined by the BF term.  In
Sec.~\ref{sec: Density operator algebra from the BF theory}, we propose
an extension of the Girvin-MacDonald-Platzman algebra to
(3+1)-dimensional spacetime by the inclusion of the vortex-density
operator in addition to the usual charge-density operator and show
that the same algebra is obeyed by massive Dirac fermions that
represent the bulk of $\mathbb{Z}^{\,}_{2}$ topological insulators in
three-dimensional space.  Finally, we close with discussions in
Sec.~\ref{sec:Discussion}.

\section{Review of Lagrangian fluids}
\label{sec:Review of Lagrangian fluids}

We begin by reviewing the Lagrangian description of fluids.%
~\cite{Jackiw04}
We consider a system of identical classical particles, 
described by coordinates $\bs{x}^{\,}_{\beta}(t)$
and velocity fields $\dot{\bs{x}}^{\,}_{\beta}(t)$,
where $\{\beta\}$ is a discrete set of particle labels.
The Lagrangian of the system reads
\begin{equation}
L\equiv
\sum_{\beta} 
\mathcal{L}
\Big(
\bs{x}^{\,}_{\beta}(t), \dot{\bs{x}}^{\,}_{\beta}(t)
\Big).
\label{eq: discrete lagrangian}
\end{equation}
For identical particles,
the choice of the particle label $\beta$ is arbitrary.
Correspondingly, the Lagrangian $L$
is invariant under any relabeling of the discrete indices 
\begin{equation}
\{ \beta \} \to \{ \beta' \}.
\label{eq: discrete symmetry}
\end{equation}
 
In the hydrodynamical description of the system, one replaces the
discrete label $\beta\in\{\beta\}$ by the label $\bs{y}\in\mathbb{R}^{3}$,
i.e., the coordinate and velocity vectors become vector fields
according to the rule
\begin{equation}
\bs{x}^{\,}_{\beta}(t)\to
\bs{x}(t,\bs{y}),
\qquad
\dot{\bs{x}}^{\,}_{\beta}(t)\to
\dot{\bs{x}}(t,\bs{y}),
\end{equation}
respectively. Here, $\bs{y}$ can be thought of as a comoving coordinate
that labels the position of an infinitesimal droplet of the fluid. 
Initially, i.e., at $t=0$, we declare that
$\bs{x}(t=0,\bs{y})=\bs{y}$. In this hydrodynamical limit,
the Lagrangian~(\ref{eq: discrete lagrangian}) becomes
\begin{equation}
L=
\int\,\mathrm{d}^{3}\bs{y}\,
\rho^{\,}_{0}\,
\mathcal{L}
\Big(
\bs{x}(t,\bs{y}), \dot{\bs{x}}(t,\bs{y})
\Big),
\label{eq:Lagrangian continuum}
\end{equation}
where the positive number $\rho^{\,}_{0}$ 
is interpreted as the mean particle density 
in $\bs{y}$-space.

The invariance of the Lagrangian~(\ref{eq: discrete lagrangian}) 
under any particle relabeling~(\ref{eq: discrete symmetry}) 
translates, in the  fluid description, 
to an emergent continuous (gauge) symmetry of the 
Lagrangian~(\ref{eq:Lagrangian continuum}) with respect
to a properly defined reparametrization
\begin{subequations}
\label{eq: continuous symmetry}
\begin{equation}
\bs{y}\to\tilde{\bs{y}}(\bs{y}).
\label{eq: continuous symmetry a}
\end{equation}
To identify this continuous symmetry, we require
that the coordinates of the fluid remain invariant, i.e.,
\begin{equation}
\tilde{\bs{x}}(t,\tilde{\bs{y}}) =  \bs{x}(t,\bs{y}),
\label{eq: continuous symmetry b}
\end{equation}
\end{subequations}
since the physical position of a particle does not depend on the 
chosen underlying parametrization. The Lagrangian%
~(\ref{eq:Lagrangian continuum}) transforms 
under the reparametrization~(\ref{eq: continuous symmetry}) as
\begin{equation}
\tilde{L}= 
\int\mathrm{d}^{3}\tilde{\bs{y}}\, 
\Big{|}\frac{\partial\bs{y}}{\partial\tilde{\bs{y}}}\Big{|}\,
\rho^{\,}_{0}\,
\mathcal{L}
\Big(
\tilde{\bs{x}}(t,\tilde{\bs{y}}), \dot{\tilde{\bs{x}}}(t,\tilde{\bs{y}})
\Big).
\label{eq: L to tilde L}
\end{equation}
Invariance of Eq.~(\ref{eq: L to tilde L}), i.e., $\tilde{L}=L$,
is then achieved provided
\begin{equation}
\Big{|} 
\frac{\partial\bs{y}}{\partial\tilde{\bs{y}}} 
\Big{|}=1.
\label{eq: def VPD condition}
\end{equation}
Condition~(\ref{eq: def VPD condition})
defines a volume preserving diffeomorphism (VPD)
if we assume that the map~(\ref{eq: continuous symmetry a})
is sufficiently smooth.

An infinitesimal VPD is defined by
\begin{subequations}
\label{eq: def VPD} 
\begin{equation}
\delta^{\,}_{\bs{f}}\,y^{\,}_{i}:=
\tilde{y}^{\,}_{i}
-
y^{\,}_{i}=
-
f^{\,}_{i}(\bm{y}),
\qquad
i=1,2,3,
\label{eq: def VPD a}
\end{equation}
where the infinitesimal vector field $\bm{f}$ must be divergence free, 
\begin{equation}
\partial^{\,}_{i}\,f^{\,}_{i}=
0,
\label{eq: def VPD b}
\end{equation}
in order to meet condition~(\ref{eq: def VPD condition}).
Here and throughout,
\begin{equation}
\partial^{\,}_{i}\equiv
\frac{\partial}{\partial y^{\,}_{i}},
\qquad
i=1,2,3.
\label{eq: def VPD c}
\end{equation}
\end{subequations} 
In three dimensional space, 
the divergence-free vector field $\bm{f}$, 
with the components $f^{\,}_{i}$
defined in Eq.~(\ref{eq: def VPD})
for $i=1,2,3$ carrying the dimension of length,
can always be parametrized (in a non-unique way) as
\begin{equation}
f^{\,}_{i}=
\epsilon^{\,}_{ijk}\,\partial^{\,}_{j}\zeta^{\,}_{k}
\end{equation}
for any smooth vector field $\bm{\zeta}$
with the components $\zeta^{\,}_{m}$
($m=1,2,3$) carrying the dimension of area. 
In the following, summation over repeated indices is 
implied and sum over the Latin indices run over the three spatial components.

In turn, the variation of the coordinate $\bm{x}$
under the transformation parametrized by $\bs{f}$ is defined by
\begin{equation}
\delta^{\,}_{\bs{f}}\,\bm{x}(t,\bm{y}):=
\tilde{\bm{x}}(t,\bm{y})
-
\bm{x}(t,\bm{y}).
\end{equation}
With the help of Eq.~(\ref{eq: continuous symmetry b})
and upon insertion of the infinitesimal transformation%
~(\ref{eq: def VPD}),
\begin{equation}
\begin{split}
\delta^{\,}_{\bs{f}}\,x^{\,}_{i}=&\,
f^{\,}_{j}\,\partial^{\,}_{j}\,x^{\,}_{i}
\\
=&\,
\epsilon^{\,}_{jlm}\,\partial^{\,}_{j}\,x^{\,}_{i}\,\partial^{\,}_{l}\,\zeta^{\,}_{m}.
\end{split}
\label{eq:change in x due to f}
\end{equation}
From the invariance of the Lagrangian%
~(\ref{eq:Lagrangian continuum}) 
under  arbitrary infinitesimal VPD defined by Eqs.%
~(\ref{eq: continuous symmetry})
and~(\ref{eq: def VPD condition}),
there follows, according to Noether's theorem, the constant of motion
\begin{subequations}
\label{eq:Noether charge 1}
\begin{equation}
\begin{split}
C^{\,}_{\bs{f}}:=&\,
\int\mathrm{d}^{3}\bs{y}\,
\rho^{\,}_{0}\,
\pi^{\,}_{i}\,\delta^{\,}_{\bs{f}}\,x^{\,}_{i}
\\
=&\,
\int\mathrm{d}^{3}\bs{y}\,
\rho^{\,}_{0}\,
\pi^{\,}_{i}\,
\epsilon^{\,}_{jlm}\,\partial^{\,}_{j}\,x^{\,}_{i}\,\partial^{\,}_{l}\,\zeta^{\,}_{m}
\\
=&\,
\int\mathrm{d}^{3}\bs{y}\,
\rho^{\,}_{0}\,
\Big(
\epsilon^{\,}_{m l j}
\partial^{\,}_{l}\,\pi^{\,}_{i}\,
\partial^{\,}_{j}\,x^{\,}_{i}\,
\Big)
\zeta^{\,}_{m},
\end{split}
\label{eq:Noether charge 1 a}
\end{equation}
where
\begin{equation}
\pi^{\,}_{i}:=
\frac
{
\partial\,\mathcal{L}
}
{
\partial\,\dot{x}^{\,}_{i}
}
\label{eq:Noether charge 1 b}
\end{equation}
\end{subequations}
is the canonical momentum and,
in deriving Eq.~(\ref{eq:Noether charge 1}), we have
made use of integration by parts and we have neglected surface terms.
Invariance of Eq.~(\ref{eq:Noether charge 1}) 
under the infinitesimal coordinate transformation%
~(\ref{eq:change in x due to f}) for an arbitrary
vector field $\bm{\zeta}$ yields 
the local conservation law
\begin{subequations}
\label{eq: noether charge}
\begin{equation}
\frac{\mathrm{d}\bm{\Lambda}}{\mathrm{d}t}=0
\end{equation}
for the vector field 
$\bm{\Lambda}$ with the components
\begin{equation}
\Lambda^{\,}_{i}:=
\epsilon^{\,}_{ijk}\,
\partial^{\,}_{j}\,\pi^{\,}_{l}\,
\partial^{\,}_{k}\,x^{\,}_{l},
\quad
i=1,2,3.
\label{eq: noether charge a}
\end{equation}
\end{subequations}
The vector field $\bm{\Lambda}$
carries the dimension of energy multiplied by time per area. 

The local density of the fluid is defined by
\begin{subequations} 
\label{eq: def rho}
\begin{equation}
\rho(t,\bm{y}):=
\rho^{\,}_{0}\,J\,
\left(\frac{\partial\bs{y}}{\partial\bs{x}}\right)(t,\bm{y}),
\end{equation}
where
\begin{equation}
J\left(\frac{\partial\bs{x}}{\partial\bs{y}}\right):=
\left|
\epsilon^{\,}_{ijk}
\frac{\partial x^{\,}_{1}}{\partial y^{\,}_{i}}
\frac{\partial x^{\,}_{2}}{\partial y^{\,}_{j}}
\frac{\partial x^{\,}_{3}}{\partial y^{\,}_k}
\right|
=
1\Big/J\left(\frac{\partial\bs{y}}{\partial\bs{x}}\right)
\end{equation}
\end{subequations}
is the Jacobian that relates the infinitesimal volume element
$\mathrm{d}^{3}\bm{y}$ to the infinitesimal volume element
$\mathrm{d}^{3}\bm{x}(t,\bm{y})$. Starting with
$\bs{x}(t=0,\bs{y})=\bs{y}$ yields an initially uniform fluid density 
$\rho(t=0,\bm{y})=\rho^{\,}_{0}$.

We define an antisymmetric two-form with the components
\begin{subequations}
\label{eq: def bij}
\begin{equation}
b^{\,}_{ij}=-b^{\,}_{ji},
\qquad i,j=1,2,3, 
\label{eq: def bij a}
\end{equation}
through
\begin{equation}
\epsilon^{\,}_{ijk}\,
b^{\,}_{jk}(t,\bs{y}):=
\rho^{\,}_{0}
\left[
x^{\,}_{i}(t,\bs{y})
-
y^{\,}_{i}
\right],
\qquad
i=1,2,3.
\label{eq: def bij b}
\end{equation}
\end{subequations}
The vector field with the components
$\epsilon^{\,}_{ijk}\,b^{\,}_{ij}$ 
carries the dimensions of inverse area and is proportional
to the deviation between the coordinate 
$x^{\,}_{i}(t,\bs{y})$ at time $t$ and its initial value $\bm{y}$
at time $t=0$.
Assuming small deviations of the fluid density away from $\rho^{\,}_{0}$, 
we may treat the two-form $b^{\,}_{ij}=-b^{\,}_{ji}$ as small. 
In terms of this field, the density of the fluid is given by
\begin{equation}
\rho=
\rho^{\,}_{0} 
-  
\epsilon^{\,}_{ijk}\,
\partial^{\,}_{i}b^{\,}_{jk}
+
\cdots,
\label{eq: density}
\end{equation}
where $\cdots$ stands for higher order terms in $b^{\,}_{ij}$.
One verifies that the transformation law
\begin{equation}
b^{\,}_{jk}
\to
b^{\,}_{jk}
+
\partial^{\,}_{j}\chi^{\,}_{k}
-
\partial^{\,}_{k}\chi^{\,}_{j}
\label{eq: gauge trafo rho}
\end{equation}
does not alter the density~\eqref{eq: density}
provided the vector field $\bm{\chi}$ 
with the components $\chi^{\,}_{i}$ for $i=1,2,3$ is smooth,
i.e., 
$\partial^{\,}_{j}\partial^{\,}_{k}\chi^{\,}_{i}=
\partial^{\,}_{k}\partial^{\,}_{j}\chi^{\,}_{i}$.
Equation~(\ref{eq: gauge trafo rho}) can also be obtained 
with the identification $\bm{\chi}=\rho^{\,}_{0}\,\bm{\zeta}/2$ from 
\begin{equation}
\epsilon^{\,}_{ijk}
\left(
\tilde{b}^{\,}_{jk}(t,\bm{y})
-
b^{\,}_{jk}(t,\bm{y})
\right):=
\rho^{\,}_{0}
\left[
\tilde{x}^{\,}_{i}(t,\bm{y})
-
x^{\,}_{i}(t,\bm{y})
\right].
\end{equation} 
Hereto, one makes use of the fact that the 2-tensor field
is antisymmetric on the left-hand side, while one makes use 
of the linearized version
of Eq.~(\ref{eq:change in x due to f}), 
whereby the approximation
$\partial^{\,}_{j}x^{\,}_{i}\approx\delta^{\,}_{ij}$
is done, on the right-hand side.
The invariance of the local density%
~(\ref{eq: density})
 under the transformation%
~(\ref{eq: gauge trafo rho}) thus
reflects the invariance of the Lagrangian%
~(\ref{eq:Lagrangian continuum}) 
under any VPD defined by Eqs.%
~(\ref{eq: continuous symmetry})
and~(\ref{eq: def VPD condition}).

We close this review of Lagrangian fluids with the
example defined by the Lagrangian
\begin{subequations}
\label{eq: def free Lagrangian continuum}
\begin{equation}
L^{\,}_{\mathrm{free}}:=
\int\,\mathrm{d}^{3}\bs{y}\,
\rho^{\,}_{0}\,
\mathcal{L}^{\,}_{\mathrm{free}}
\label{eq: def free Lagrangian continuum a}
\end{equation}
with the local Lagrangian
\begin{equation}
\mathcal{L}^{\,}_{\mathrm{free}}:=
\frac{m}{2}\,
\dot{\bm{x}}^{2}.
\label{eq: def free Lagrangian continuum b}
\end{equation}
\end{subequations}
This Lagrangian describes a fluid of non-interacting
and identical classical particles of mass $m$.
The canonical momentum~(\ref{eq:Noether charge 1})
becomes the usual impulsion
\begin{equation}
\bm{\pi}=
m\,
\dot{\bm{x}}.
\end{equation}
The local conserved vector field~(\ref{eq: noether charge a})
becomes
\begin{equation}
\Lambda^{\,}_{i}=
m\,
\epsilon^{\,}_{ijk}\,
\partial^{\,}_{j}\,
\dot{x}^{\,}_{l}\,
\partial^{\,}_{k}\,
x^{\,}_{l},
\label{eq: vortex helicity for free fluid}
\end{equation}
whose conserved integral is called the vortex helicity
and is related to a Chern number 
(see Ref.~\onlinecite{Jackiw04}).
In terms of the two-form defined in Eq.%
~(\ref{eq: def bij b}),
the canonical momentum is (exactly) given by
\begin{equation}
\pi^{\,}_{i}=
\frac{m}{\rho^{\,}_{0}}\,
\epsilon_{ijk}\,
\dot{b}^{\,}_{jk},
\end{equation}
while the vortex helicity~(\ref{eq: vortex helicity for free fluid})
is given by
\begin{equation}
\Lambda^{\,}_{i}=
2\,
\frac{m}{\rho^{\,}_{0}}\,
\partial^{\,}_{j}
\dot{b}^{\,}_{ij}
+
\cdots
\end{equation}
to leading order in powers of the two-form defined in Eq.%
~(\ref{eq: def bij b}).

\section{BF Lagrangian for an incompressible polar fluid}
\label{sec:BF Lagrangian for an incompressible polar fluid}

\subsection{Definition}

We start from the discrete set $\{\beta\}$ that labels
identical particles with a mass $m$.
We associate to any label $\beta$ the coordinate
$\bm{x}^{\,}_{\beta}(t)$,
the velocity
$\dot{\bm{x}}^{\,}_{\beta}(t)$,
and the polar vector
$\bm{d}^{\,}_{\beta}(t)$
whose dimension we choose for later convenience to be that of an inverse length.
We then endow a Lagrangian dynamics to these degrees of freedom
by defining
\begin{subequations}
\label{eq: discrete BF lagrangian}
\begin{equation}
L^{\,}_{\mathrm{pol}}:=
\sum_{\beta}
\mathcal{L}^{\,}_{\mathrm{pol}}
\Big(
\dot{\bs{x}}^{\,}_{\beta}(t),\bs{d}^{\,}_{\beta}(t)
\Big)
\label{eq: discrete Pol lagrangian a}
\end{equation}
where
\begin{equation}
\mathcal{L}^{\,}_{\mathrm{pol}}
\Big(
\dot{\bs{x}}^{\,}_{\beta}(t),\bs{d}^{\,}_{\beta}(t)
\Big):=
-\frac{g}{2\pi}\,
\bs{d}^{\,}_{\beta}(t)
\cdot
\dot{\bs{x}}^{\,}_{\beta}(t).
\label{eq: discrete Pol lagrangian b}
\end{equation}
\end{subequations} 
The real-valued coupling $g$ 
carries the dimension of energy multiplied by time.
The multiplicative factor $(-1)/(2\pi)$ is chosen for later convenience.

The hydrodynamic limit of
the Lagrangian~(\ref{eq: discrete BF lagrangian})
is the Lagrangian polar fluid
\begin{subequations}
\label{eq: Pol lagrangian continuum}
\begin{equation}
L^{\,}_{\mathrm{pol}}:=
\int\,\mathrm{d}^{3}\bs{y}\,
\rho^{\,}_{0}\,
\mathcal{L}^{\,}_{\mathrm{pol}},
\label{eq: Pol Lagrangian continuum a}
\end{equation}
with the local Lagrangian
\begin{equation}
\mathcal{L}^{\,}_{\mathrm{pol}}:=
-\frac{g}{2\pi}\,
\bs{d}
\cdot
\dot{\bs{x}}
\label{eq: Pol lagrangian continuum b}
\end{equation}
\end{subequations}
carrying the dimension of energy, for
the positive number $\rho^{\,}_{0}$ 
is again interpreted as the mean particle density 
in $\bs{y}$-space.

The Lagrangian density~(\ref{eq: Pol lagrangian continuum b}) is
invariant under simultaneous rotations of the coordinate and polar
vectors. Moreover, it is the unique scalar that is linear in both
$\bm{d}$ and $\bs{x}$ and of first order in the time derivative, up to
a total time derivative. In addition to the rotational symmetry, two
discrete symmetries are notable. The first is parity, 
\begin{equation}
\mathcal{P}:\quad
\begin{cases}
\bs{d}(t,\bm{y})\to
-
\bs{d}(t,\bm{y}),
\\
\bs{x}(t,\bm{y})\to
-\bs{x}(t,\bm{y}),
\\
\dot{\bs{x}}(t,\bm{y})\to
-\dot{\bs{x}}(t,\bm{y}).
\end{cases}
\label{eq: def polar trsf}
\end{equation}  
The second is time-reversal symmetry
\begin{equation}
\mathcal{T}^{\,}_{\pm}:\quad
\begin{cases}
\bs{d}(t,\bm{y})\to
\pm
\bs{d}(-t,\bm{y}),
\\
\bs{x}(t,\bm{y})\to
+\bs{x}(-t,\bm{y}),
\\
\dot{\bs{x}}(t,\bm{y})\to
-\dot{\bs{x}}(-t,\bm{y}),
\end{cases}
\label{eq: def TRS trsf}
\end{equation}
where the $\pm$ sign choice depends on the nature of the dipoles. 
It is $+$ for electric dipoles, while it is $-$ for magnetic dipoles. 
The Lagrangian density~(\ref{eq: Pol lagrangian continuum b}) is invariant
under $\mathcal{P}$ and under $\mathcal{T}^{\,}_{-}$ 
(applicable to magnetic moments). Most importantly,
the polar fluid is invariant under any VPD defined by Eqs.%
~(\ref{eq: continuous symmetry}) and~(\ref{eq: def VPD condition}). We
focus primarily on the invariance under VPD.

We are after the local density
\begin{subequations}
\label{eq: def two-fluid density + vorticity}
\begin{equation}
\rho(t,\bs{y}):=
\rho^{\,}_{0}\,
J\left(\frac{\partial\bs{y}}{\partial\bs{x}}\right)
\label{eq: def two-fluid density + vorticity a}
\end{equation}
and the local conserved Noether vorticity field 
$\bm{\Lambda}$ with the components
\begin{equation}
\begin{split}
\Lambda^{\,}_{i}:=&\,
\epsilon^{\,}_{ijk}\,
\partial^{\,}_{j}\,\pi^{\,}_{l}\,
\partial^{\,}_{k}\,x^{\,}_{l}
\\
=&\,
-\frac{g}{2\pi}\,
\epsilon^{\,}_{ijk}
\frac{\partial d^{\,}_{l}}{\partial y^{\,}_{j}}
\frac{\partial x^{\,}_{l}}{\partial y^{\,}_{k}}
\end{split}
\label{eq: def two-fluid density + vorticity b}
\end{equation}
\end{subequations}
for $i=1,2,3$.
The density is even under either the transformation%
~(\ref{eq: def polar trsf})
or the transformation%
~(\ref{eq: def TRS trsf}).
The vortex helicity is odd under either the transformation%
~(\ref{eq: def polar trsf})
or the transformation%
~(\ref{eq: def TRS trsf}).

We parametrize the coordinates 
$x^{\,}_{1},x^{\,}_{2},x^{\,}_{3}$ according to
\begin{equation}
x^{\,}_{i}(t,\bs{y})=:
y^{\,}_{i} 
+
\frac{1}{\rho^{\,}_{0}}\,
\epsilon^{\,}_{ijk}\,b^{\,}_{jk}(t,\bs{y}).
\end{equation}
As was the case with Eq.~(\ref{eq: def bij b}),
the antisymmetric two-form with the components
$b^{\,}_{jk}(t,\bs{y})=-b^{\,}_{kj}(t,\bs{y})$
encodes, up to a contraction with $(1/\rho^{\,}_{0})\,\epsilon^{\,}_{ijk}$,
the deviation between the comoving coordinate $\bm{y}$
and the coordinate $\bm{x}^{\,}(t,\bm{y})$ at time $t$ 
in the polar fluid.

Under the assumptions that both $b^{\,}_{ij}$ and $\bm{d}$ 
remain small for all times and for all comoving coordinates, 
one finds the relations
\begin{subequations}
\label{eq: linearized density + vorticity}
\begin{equation}
\rho(t,\bs{y})=
\rho^{\,}_{0}
-
\epsilon^{\,}_{ijk}\,
\partial_{i}b^{\,}_{jk}(t,\bs{y})
+
\cdots,
\label{eq: linearized density + vorticity a}
\end{equation}
and
\begin{equation}
\Lambda^{\,}_{i}(t,\bs{y})=
-\frac{g}{2\pi}\,
\epsilon^{\,}_{ijk}\,
\partial^{\,}_{j}d^{\,}_{k}(t,\bs{y})
+
\cdots,
\label{eq: linearized density + vorticity b}
\end{equation}
\end{subequations}
to linear order in the fields $b^{\,}_{ij}$ and $d^{\,}_{i}$, 
for the local density~(\ref{eq: def two-fluid density + vorticity a})
and local vortex helicity~(\ref{eq: def two-fluid density + vorticity b}), 
respectively.

Equation~(\ref{eq: linearized density + vorticity a})
is invariant under the transformation
\begin{subequations}
\label{eq: linearized VPD}
\begin{equation}
b^{\,}_{jk}\to 
b^{\,}_{jk}
+
\partial^{\,}_{j}\chi^{\,}_{k}
-
\partial^{\,}_{k}\chi^{\,}_{j}
\label{eq: linearized VPD a}
\end{equation}
for any smooth vector field $\bm{\chi}$.
Equation~(\ref{eq: linearized density + vorticity b})
is invariant under
\begin{equation}
d^{\,}_{k}\to 
d^{\,}_{k}
+
\partial^{\,}_{k}\xi,
\label{eq: linearized VPD b}
\end{equation}
\end{subequations}
for any smooth scalar field $\xi$.
The linearized local density
(\ref{eq: linearized density + vorticity a}) 
is even under either the transformation%
~(\ref{eq: def polar trsf})
or the transformation%
~(\ref{eq: def TRS trsf}).
The linearized local vortex helicity
(\ref{eq: linearized density + vorticity b}) 
is odd under either the transformation%
~(\ref{eq: def polar trsf})
or the transformation%
~(\ref{eq: def TRS trsf}).

The local Lagrangian~(\ref{eq: Pol lagrangian continuum}),
takes the linearized form (up to total derivatives)
\begin{equation}
\mathcal{L}^{\,}_{\mathrm{pol}}=
\frac{g}{2\pi\,\rho^{\,}_{0}}\,
\epsilon^{\,}_{ijk}\,
\dot{d}^{\,}_{i}\,
b^{\,}_{jk},
\label{eq: linearized polar lag}
\end{equation}
where we recall that 
$g$,
$\rho^{\,}_{0}$,
$d^{\,}_{i}$,
and
$b^{\,}_{jk}$
carry the dimensions of
energy multiplied by time,
inverse volume,
inverse length, 
and inverse area, respectively.

A VPD defined by Eqs.\ (\ref{eq: continuous symmetry}) and (\ref{eq:
  def VPD condition}) leaves the local density%
~(\ref{eq: def two-fluid density + vorticity a}) of the polar fluid
invariant. This symmetry is realized by the symmetry under the
transformation%
~(\ref{eq: linearized VPD a}) of the linearized local density%
~(\ref{eq: linearized density + vorticity a}) and must hold at the
level of the linearized local Lagrangian
(\ref{eq: linearized polar lag}). 
Indeed it does, as we now verify. The transformation
law of $\mathcal{L}^{\,}_{\mathrm{pol}}$ under the infinitesimal VPD
(\ref{eq: linearized VPD a}) is
\begin{equation}
\mathcal{L}^{\,}_{\mathrm{pol}}\to
\mathcal{L}^{\,}_{\mathrm{pol}}
+
2\times
\frac{g}{2\pi\,\rho^{\,}_{0}}\,
\dot{\bs{d}}\,
\cdot
\left(
\bs{\nabla}
\wedge
\bs{\chi}
\right).
\end{equation}
Since the vector field $\bs{\chi}$ is arbitrary, to enforce the
symmetry under VPD we must demand that
\begin{equation}
\frac{\mathrm{d}}{\mathrm{d}t}
\left(
\bs{\nabla}
\wedge
\bs{d}
\right)=0.
\label{eq: linearized VPD}
\end{equation}
Now, Eq.~(\ref{eq: linearized VPD}) 
follows from 
\begin{equation}
\frac{
\mathrm{d}\bs{\Lambda}
     }
     {
\mathrm{d}t
     }=0,
\end{equation} 
to linear order, as can be observed from
Eq.%
~(\ref{eq: linearized density + vorticity b}).
[As we did to reach Eq.~(\ref{eq: noether charge}),
we are ignoring boundary terms when performing
partial integrations.]

The linearized local Lagrangian~(\ref{eq: linearized polar lag})
is proportional to the Lagrangian
density of the topological BF field theory defined by
Eq.~(\ref{eq: def BF L})
in the temporal gauge defined by the conditions
\begin{equation}
d^{\,}_{0}=0,
\qquad
b^{\,}_{0i}=0,
\qquad
i=1,2,3.
\end{equation}
A BF field theory is an example of a topological field theory.
Topological field theories are interpreted in physics
as effective descriptions at long distances, low energies,
and vanishing temperature
of quantum Hamiltonians with spectral gaps separating
the ground state manifolds from all excited states. This
observation motivates the following definition.
The VPD polar fluid is said to be incompressible if it 
has the constant density
\begin{equation}
\rho=\rho^{\,}_{0} \;.
\end{equation}

Without loss of generality, we consider henceforth a magnetic dipolar
fluid, in which any non-vanishing value taken by the conserved quantity
$\bs{\Lambda}$ breaks the symmetry under $\mathcal{T}^{\,}_{-}$ 
defined in Eq.\
(\ref{eq: def TRS trsf}).
We say that the VPD polar fluid is time-reversal symmetric if and only if
\begin{equation}
\bs{\Lambda}=0.
\end{equation}
[The same conclusion is reached for an electric polar fluid,
in which case it is the symmetry under 
$\mathcal{P}$ defined in Eq.\
(\ref{eq: def polar trsf}) that implies 
$\bs{\Lambda}=0$.]

Incompressibility of a time-reversal symmetric (magnetic) polar fluid
is automatically implemented with the help of the Lorentz covariant
extension of $\mathcal{L}^{\,}_{\mathrm{pol}}$ given by (we set the
speed of light $c$ to be unity, $c=1$,
and
$\mu,\nu,\lambda,\sigma=0,1,2,3$)
\begin{equation}
S^{\,}_{\mathrm{BF}}:=
\int\mathrm{d}^{4}y\,
\mathcal{L}^{\,}_{\mathrm{BF}},
\qquad
\mathcal{L}^{\,}_{\mathrm{BF}}:=
\frac{g}{2\pi}\,
\epsilon^{\mu\nu\lambda\sigma}\,
b^{\,}_{\mu\nu}\,
\partial^{\,}_{\lambda}
d^{\,}_{\sigma}.
\label{eq: def BF L}
\end{equation}
Indeed, the equations of motion that follow from
$\mathcal{L}^{\,}_{\mathrm{BF}}$ 
are the conservation laws for the matter current
\begin{equation} 
j^{\mu}:=
\frac{1}{2\pi}\,
\epsilon^{\mu\nu\lambda\sigma}\,
\partial^{\,}_{\nu}\, 
b^{\,}_{\lambda\sigma},
\qquad
\partial^{\,}_{\mu}j^{\mu}=0,
\label{eq: conserved jmu current}
\end{equation}
and for the vortex-helicity currents
\begin{equation}
J^{\mu\nu}:=
\frac{1}{2\pi}\,
\epsilon^{\mu\nu\lambda\sigma}\,
\partial^{\,}_{\lambda}d^{\,}_{\sigma},
\qquad
\partial^{\,}_{\mu}J^{\mu\nu}=0.
\label{eq: conserved Jmunu current}
\end{equation}
The time-component 
\begin{equation}
j^{0}=
\frac{1}{2\pi}\,
\epsilon^{ijk}\,
\partial^{\,}_{i}b^{\,}_{jk}=
\frac{1}{2\pi}\,
\epsilon^{\,}_{ijk}\,
\partial^{\,}_{i}b^{\,}_{jk}
\end{equation}
of the one-form $j^{\mu}$ is 
the density $(\rho^{\,}_{0}-\rho)/2\pi$ from Eq.%
~(\ref{eq: linearized density + vorticity a}).
The time-component
\begin{equation} 
J^{0i}=
\frac{1}{2\pi}\,
\epsilon^{ijk}\,
\partial^{\,}_{j}d^{\,}_{k}=
\frac{1}{2\pi}\,
\epsilon^{\,}_{ijk}\,
\partial^{\,}_{j}d^{\,}_{k}
\end{equation}
of the two-form $J^{\mu\nu}$ 
defines the vortex helicity $-\bs{\Lambda}/g$, see Eq.%
~(\ref{eq: linearized density + vorticity b}).
The difference between the Lagrangian density
(\ref{eq: linearized polar lag})
and its Lorentz covariant extension~(\ref{eq: def BF L}) 
is that the latter contains terms of the form
$d^{\,}_{0}\,\epsilon^{\,}_{ijk}\,\partial^{\,}_{i}b^{\,}_{jk}/(2\pi)$
and 
$-b^{\,}_{0i}\,\epsilon^{\,}_{ijk}\,\partial^{\,}_{j}d^{\,}_{k}/(2\pi)$,
which, upon using Eqs.~(\ref{eq: linearized density + vorticity a})
and (\ref{eq: linearized density + vorticity b}),
are rewritten as
\begin{subequations}
\label{eq: Lagrange multiplyers in 2+1 BF}
\begin{equation}
\frac{1}{2\pi}\,
d^{\,}_{0}\,\epsilon^{\,}_{ijk}\,\partial^{\,}_{i}b^{\,}_{jk}=
\frac{1}{2\pi}\,
d^{\,}_{0}\,
\left(\rho^{\,}_{0}-\rho\right)
\label{eq: Lagrange multiplyers in 2+1 BF a}
\end{equation} 
and
\begin{equation}
\frac{g}{2\pi}\,
b^{\,}_{0i}\,\epsilon^{\,}_{ijk}\,\partial^{\,}_{j}d^{\,}_{k}=
b^{\,}_{0i}\,\Lambda^{\,}_{i},
\label{eq: Lagrange multiplyers in 2+1 BF b}
\end{equation} 
\end{subequations}
respectively.
Upon quantization of the theory, say by defining the path integral
\begin{equation}
Z^{\,}_{\mathrm{BF}}:=
\int\mathcal{D}[d,b]\,
e^{+\mathrm{i}S^{\,}_{\mathrm{BF}}/\hbar},
\label{eq: partition fct BF}
\end{equation}
the fields $d^{\,}_{0}$ and $b^{\,}_{0i}$ 
take the role of Lagrange multipliers that enforce that
the ground state has the constant density $\rho=\rho^{\,}_{0}$ and
the vanishing vortex helicity $\bs{\Lambda}=0$
as a consequence of 
Eqs.~(\ref{eq: Lagrange multiplyers in 2+1 BF a})
and
(\ref{eq: Lagrange multiplyers in 2+1 BF b}),
respectively. The vanishing vortex helicity $\bs{\Lambda}=0$
automatically enforces the weaker condition
$\mathrm{d}\bs{\Lambda}/\mathrm{d}t=0$
that any VPD-symmetric polar fluid must fulfill.

The assumption that both $\bm{d}$ and $b^{\,}_{ij}$ 
remain small is self-consistent, for the equal-time and local
expectation values
\begin{equation}
\left\langle
d^{2}_{i}(t,\bm{y})
\right\rangle^{\,}_{\mathrm{BF}}\propto I
\qquad
\left\langle
b^{2}_{ij}(t,\bm{y})
\right\rangle^{\,}_{\mathrm{BF}}\propto I,
\end{equation}
for any $i,j=1,2,3$ are proportional to the integral
\begin{equation}
I:=
\int\limits_{0}^{1/\mathfrak{a}}\,
\mathrm{d}^{3}\bs{k}\,
\frac{1}{|\bs{k}|}
\propto
\left(\frac{1}{\mathfrak{a}}\right)^{2}.
\end{equation}
Here, $\mathfrak{a}$ is a short-distance cutoff below which
the hydrodynamical approximation is meaningless.

We close this discussion of a VPD, incompressible,
and time-reversal symmetric polar fluid by observing that
it is perfectly legitimate to add a term like $\bs{d}^{2}$ 
to the BF action, thereby breaking the independence on the
metric, Lorentz covariance, and
the $U(1)$ gauge symmetry associated to the $\bs{d}$ field.
The $U(1)$ gauge symmetry associated to the $\bs{d}$ field
is a mere signature for the fact that
the vortex helicity is the rotation of the $\bs{d}$ field.
On the other hand, the VPD symmetry, 
which is represented by the symmetry 
of the BF action~(\ref{eq: linearized polar lag})
under the transformation (\ref{eq: linearized VPD a}),
must be preserved to any order in a gradient expansion. 

\subsection{Coupling the conserved currents to sources}

The local conservation laws%
~(\ref{eq: conserved jmu current})
and%
~(\ref{eq: conserved Jmunu current})
suggest that we attribute to the coordinate $\bs{x}(t,\bs{y})$
the conserved electric charge $e$ and that we attribute to the polar
vector $\bs{d}(t,\bs{y})$ the conserved vortex charge $s$.
Correspondingly, we may interpret the one form
$A^{\mu}$ and the antisymmetric two form $B^{\mu\nu}=-B^{\nu\mu}$ 
entering the Lagrangian density
\begin{equation}
\begin{split}
\mathcal{L}^{\,}_{\mathrm{ext}}:=&\,
e\,
j^{\mu}\,
A^{\,}_{\mu}
+
s\,
J^{\mu\nu}\,
B^{\,}_{\mu\nu}
\\
=&\,
\frac{e}{2\pi}\,
\epsilon^{\mu\nu\lambda\sigma}\,
\partial^{\,}_{\nu}b^{\,}_{\lambda\sigma}\,
A^{\,}_{\mu}
+
\frac{s}{2\pi}\,
\epsilon^{\mu\nu\lambda\sigma}\,
\partial^{\,}_{\lambda}
d^{\,}_{\sigma}\,
B^{\,}_{\mu\nu}
\end{split}
\label{eq: def e and s}
\end{equation}
as the source fields needed to generate all the correlation functions
for the conserved currents $j^{\mu}$ and $J^{\mu\nu}$ from the
BF theory defined by Eqs.~(\ref{eq: partition fct BF})
and~(\ref{eq: def BF L}), 
respectively. If we assign $A^{\,}_{\mu}$ and $B^{\,}_{\mu\nu}$
the dimensions of inverse length and inverse area, respectively,
then the couplings $e$ and $s$ carry the dimensions of energy
multiplied by length.

If we ignore total derivatives, the equations of motion obeyed by
$\mathcal{L}^{\,}_{\mathrm{BF}}+\mathcal{L}^{\,}_{\mathrm{ext}}$
upon variation with respect to $b^{\,}_{\mu\nu}$
for fixed $\mu,\nu=0,1,2,3$ are
\begin{equation}
0=
\epsilon^{\mu\nu\lambda\sigma}
\left(
g\,
\partial^{\,}_{\lambda}d^{\,}_{\sigma}
+
e\,
\partial^{\,}_{\lambda}A^{\,}_{\sigma}
\right).
\end{equation}
If we introduce the antisymmetric two forms
\begin{equation}
f^{\,}_{\lambda\sigma}:=
\partial^{\,}_{\lambda}d^{\,}_{\sigma}
-
\partial^{\,}_{\sigma}d^{\,}_{\lambda},
\qquad
F^{\,}_{\lambda\sigma}:=
\partial^{\,}_{\lambda}A^{\,}_{\sigma}
-
\partial^{\,}_{\sigma}A^{\,}_{\lambda},
\end{equation}
for some given $\lambda,\sigma=0,1,2,3$,
we may write the equations of motion obeyed by
$\mathcal{L}^{\,}_{\mathrm{BF}}+\mathcal{L}^{\,}_{\mathrm{ext}}$
upon variation with respect to $b^{\,}_{\mu\nu}$
for fixed $\mu,\nu=0,1,2,3$ as
\begin{equation}
f^{\,}_{\lambda\sigma}=
-
\frac{e}{g}\,
F^{\,}_{\lambda\sigma}.
\label{eq: equations motion from variation by bmunu}
\end{equation}
We interpret $F^{\,}_{\mu\nu}$ 
as the field strengths in electromagnetism, i.e.,
\begin{equation}
E^{\,}_{i}:=
\partial^{\,}_{0} A^{\,}_{i}-\partial^{\,}_{i}A^{\,}_{0},
\qquad i=1,2,3,
\end{equation}
are the three components of the electric field $\bs{E}$ and
\begin{equation}
B^{\,}_{i}:=
\epsilon^{\,}_{ijk}\partial^{\,}_{j}A^{\,}_{k},
\qquad i=1,2,3,
\end{equation}
are the three components of the magnetic field $\bs{B}$.
The equations of motion%
~({\ref{eq: equations motion from variation by bmunu})
bind the electromagnetic-like field strength of the polar four vector $d^{\mu}$ 
to the external electromagnetic field according to the rule
\begin{equation}
E^{\,}_{i}=
-
\frac{g}{e}
\left(
\partial^{\,}_{0}d^{\,}_{i}-\partial^{\,}_{i}d^{\,}_{0}
\right),
\qquad i=1,2,3,
\end{equation}
and
\begin{equation}
B^{\,}_{i}=
-
\frac{g}{e}\,
\epsilon^{\,}_{ijk}\partial^{\,}_{j}d^{\,}_{k},
\qquad i=1,2,3.
\end{equation}
This parallels the picture of the (fractional)
quantum Hall effect where (fractionally) charged excitations are bound
to magnetic flux quanta.
The homogeneous Maxwell equations (in units with the speed of light
$c=1$) 
\begin{subequations}
\begin{equation}
\bs{\nabla}\cdot\bs{B}=0,
\qquad
\bs{\nabla}\wedge\bs{E}+\dot{\bs{B}}=0,
\end{equation}
are automatically satisfied as a consequence of the Bianchi identity
\begin{equation}
\mathcal{F}^{\mu\nu}:=\frac{1}{2}\epsilon^{\mu\mu\lambda\sigma}\,F^{\,}_{\lambda\sigma}
\Longrightarrow
\partial^{\,}_{\mu}\mathcal{F}^{\mu\nu}=0.
\end{equation}
\end{subequations}
With the help of the equations of motion%
~(\ref{eq: equations motion from variation by bmunu}),
the vortex helicity 
\begin{equation}
\Lambda^{\,}_{i}:=
-
\frac{g}{2\pi}\,
\epsilon^{\,}_{ijk}\,
\partial^{\,}_{j}d^{\,}_{k}=
\frac{e}{2\pi}\,B^{\,}_{i},
\qquad i=1,2,3,
\end{equation}
must then obey the homogeneous differential equations
\begin{equation}
\bs{\nabla}\cdot\bs{\Lambda}=0,
\qquad
\bs{\nabla}\wedge\bs{E}
+
\frac{2\pi}{e}\,\dot{\bs{\Lambda}}=0.
\end{equation}

The equations of motion obeyed by
$\mathcal{L}^{\,}_{\mathrm{BF}}+\mathcal{L}^{\,}_{\mathrm{ext}}$
upon variation with respect to 
$d^{\,}_{\sigma}$
for fixed $\sigma=0,1,2,3$ are
\begin{equation}
0=
\epsilon^{\mu\nu\lambda\sigma}\,
\partial^{\,}_{\lambda}
\left(
g\,
b^{\,}_{\mu\nu}
+
s\,
B^{\,}_{\mu\nu}
\right).
\end{equation}

\subsection{Quadratic order in the gradient expansion}
\label{subsec: Quadratic order in the gradient expansion}

The Lagrangian density 
$\mathcal{L}^{\,}_{\mathrm{BF}}+\mathcal{L}^{\,}_{\mathrm{ext}}$ 
is of first order in a gradient expansion.
To second order in a gradient expansion,
the local extensions to
$\mathcal{L}^{\,}_{\mathrm{BF}}+\mathcal{L}^{\,}_{\mathrm{ext}}$ 
that are Lorentz scalars or pseudoscalars are the following.

There is the Thirring current-current interaction
\begin{equation}
\begin{split}
\mathcal{L}^{\,}_{\mathrm{Th}}:=&\,
g^{\,}_{\mathrm{Th}}\,
j^{\,}_{\mu}\,j^{\mu}
\\
=&\,
g^{\,}_{\mathrm{Th}}\,
\delta^{\nu\lambda\sigma}_{\nu'\lambda'\sigma'}\,
\partial^{\,}_{\nu}b^{\,}_{\lambda\sigma}\,
\partial^{\nu'}b^{\lambda'\sigma'},
\end{split}
\end{equation}
where $\delta^{\mu\lambda\sigma}_{\mu'\lambda'\sigma'}$
is a generalized Kroenecker symbol,
the conserved current $j^{\,}_{\mu}$ is defined
in Eq.~(\ref{eq: conserved jmu current}),
and the real-valued coupling $g^{\,}_{\mathrm{Th}}$
carries the dimension of energy multiplied by time and area.

There is the Maxwell term
\begin{equation}
\begin{split}
\mathcal{L}^{\,}_{\mathrm{Ma}}:=&\,
g^{\,}_{\mathrm{Ma}}\,
J^{\,}_{\mu\nu}\,J^{\mu\nu}
\\
=&\,
2\,
g^{\,}_{\mathrm{Ma}}\,
\delta^{\lambda\sigma}_{\lambda'\sigma'}\,
\partial^{\,}_{\lambda}d^{\,}_{\sigma}\,
\partial^{\lambda'}d^{\sigma'},
\end{split}
\end{equation}
where $\delta^{\lambda\sigma}_{\lambda'\sigma'}$
is a generalized Kroenecker symbol,
the conserved current $J^{\,}_{\mu\nu}$ is defined
in Eq.~(\ref{eq: conserved Jmunu current}),
and the real-valued coupling $g^{\,}_{\mathrm{Ma}}$
carries the dimension of energy multiplied by time.

Finally, there is the pseudoscalar
\begin{equation}
\mathcal{L}^{\,}_{\theta}:=
\frac{\theta}{8\pi^{2}}\,
\epsilon^{\mu\nu\lambda\sigma}
\partial^{\,}_{\mu}d^{\,}_{\nu}
\partial^{\,}_{\lambda}d^{\,}_{\sigma},
\end{equation}
where the real-valued $\theta$ carries the dimension of energy multiplied 
by time. This is a the topological axion term, a total derivative for smooth
configurations of the field $d^{\,}_{\mu}$. Singular points at which 
$d^{\,}_{\mu}$ is multivalued are sources for $\bs{\Lambda}$ 
(magnetic monopoles). 
Due to the Witten effect,~\cite{Witten79,Hughes10,Rosenberg10} 
such a point source for $\bs{\Lambda}$ carries a point charge 
$q=\theta\,e/(2\,\pi)$.

\subsection{Topological excitations}

The VPD, incompressible, and time-reversal symmetric polar fluid 
governed by Eqs.\
(\ref{eq: partition fct BF})
and
(\ref{eq: def BF L})
is described by a BF topological field theory. 
It supports static excitations bound to
point and line singularities as we now show.

We consider the static parametrization of the polar incompressible
fluid defined by the map
\begin{equation}
\bs{x}(\bs{y}):=
f(\bs{y}^{2})\,
\bs{y},
\label{eq: diffeomorphism outside disc}
\end{equation}
which we require to be diffeomorphic almost everywhere.
The real-valued $f$ is not arbitrary, 
for we demand that the Jacobian
\begin{subequations}
\label{eq: Jacobian is unity}
\begin{equation}
J\left(\frac{\partial\bs{y}}{\partial\bs{x}}\right)=1,
\label{eq: Jacobian is unity a}
\end{equation}
i.e., we interpret the map $\bs{y}\mapsto\bs{x}(\bs{y})$ as a VPD
almost everywhere. In this way,
\begin{equation}
\rho(\bs{y})=
\rho^{\,}_{0}\,
J\left(\frac{\partial\bs{y}}{\partial\bs{x}}\right)=
\rho^{\,}_{0}
\label{eq: Jacobian is unity b}
\end{equation}
\end{subequations}
almost everywhere [recall Eq.~(\ref{eq: def rho})].
Condition~(\ref{eq: Jacobian is unity}) amounts to
solving the non-linear differential equation
\begin{equation}
f^{3}+2\,f'\,f^{2}\,\bs{y}^{2}=1,
\qquad
f':=\frac{\mathrm{d}f}{\mathrm{d}\bs{y}^{2}}.
\label{eq: differential equation for f}
\end{equation}
Solutions to the differential equations%
~(\ref{eq: differential equation for f})
are of the form
\begin{equation}
f(y):=
\left(
1
\pm
\frac{c^{3}}{y^{3}}
\right)^{1/3},
\end{equation}
where $\pm\ln c^{2}$ is a real-valued integration constant.
Admissible real-valued solutions of the form%
~(\ref{eq: diffeomorphism outside disc})
must satisfy simultaneously
\begin{subequations}
\label{eq: solution diffeomorphism outside disc}
\begin{equation}
\bs{x}(\bs{y})=
\bs{y}\,
\left(
1
\pm
\frac{{r}^{3}_{e}}{|\bs{y}|^{3}}
\right)^{1/3}
\end{equation}
and
\begin{equation}
\bs{y}(\bs{x})=
\bs{x}\,
\left(
1
\mp
\frac{{r}^{3}_{e}}{|\bs{x}|^{3}}
\right)^{1/3},
\end{equation}
\end{subequations}
i.e., either ${r}_{e}\leq|\bs{x}|$ 
if the sign $+\ln c^{2}$ is chosen for the integration constant
or ${r}_{e}\leq|\bs{y}|$ 
if the sign $-\ln c^{2}$ is chosen for the integration constant.

Figure~\ref{fig: puncture} illustrates
the fact that the fluid is excluded within a radius ${r}^{\,}_{e}$
by the almost everywhere diffeomorphic map
(\ref{eq: solution diffeomorphism outside disc}). 
This excluded volume can be interpreted as a hole of total particle number
\begin{equation}
q^{\,}_{e}:=
\rho^{\,}_{0}\;\frac{4\pi}{3}{r}^{3}_{e}
\label{eq: def puncture size}
\end{equation}
At distances from the origin that are much larger than 
$r^{\,}_{e}$,
say $|\bs{y}|\gg r^{\,}_{e}$, the linear approximation%
~\eqref{eq: def bij} is valid and yields the long-distance behavior
\begin{equation}
b^{\,}_{jk}\sim
\frac{q^{\,}_{e}}{8\pi}\,
\epsilon^{\,}_{jki}\,
\frac{y^{\,}_{i}}{|\bs{y}|^{3}}.
\label{eq: monopole charge qe}
\end{equation}

A second type of topological defect of
a VPD, incompressible, and time-reversal symmetric polar fluid 
consists in allowing the vortex helicity field $\bm{\Lambda}$
not to be divergence free along a string. 
A static line defects comes in the form of an infinitesimally thin
solenoid. A flux tube carrying the 
dimensionless flux $q^{\,}_{s}$ 
that runs through the origin along
the $y^{\,}_{3}$-axis obeys the asymptotics
\begin{equation}
d^{\,}_{1}\sim
+\frac{q^{\,}_{s}}{2\pi}\,
\frac{y^{\,}_{2}}{y^{2}_{1}+y^{2}_{2}},
\quad
d^{\,}_{2}\sim
-
\frac{q^{\,}_{s}}{2\pi}\,
\frac{y^{\,}_{1}}{y^{2}_{1}+y^{2}_{2}},
\quad
d^{\,}_{3}\sim
0.
\label{eq: magnetic flux phim}
\end{equation}

\begin{figure}[t]
\centering
\includegraphics[width=0.4\textwidth]{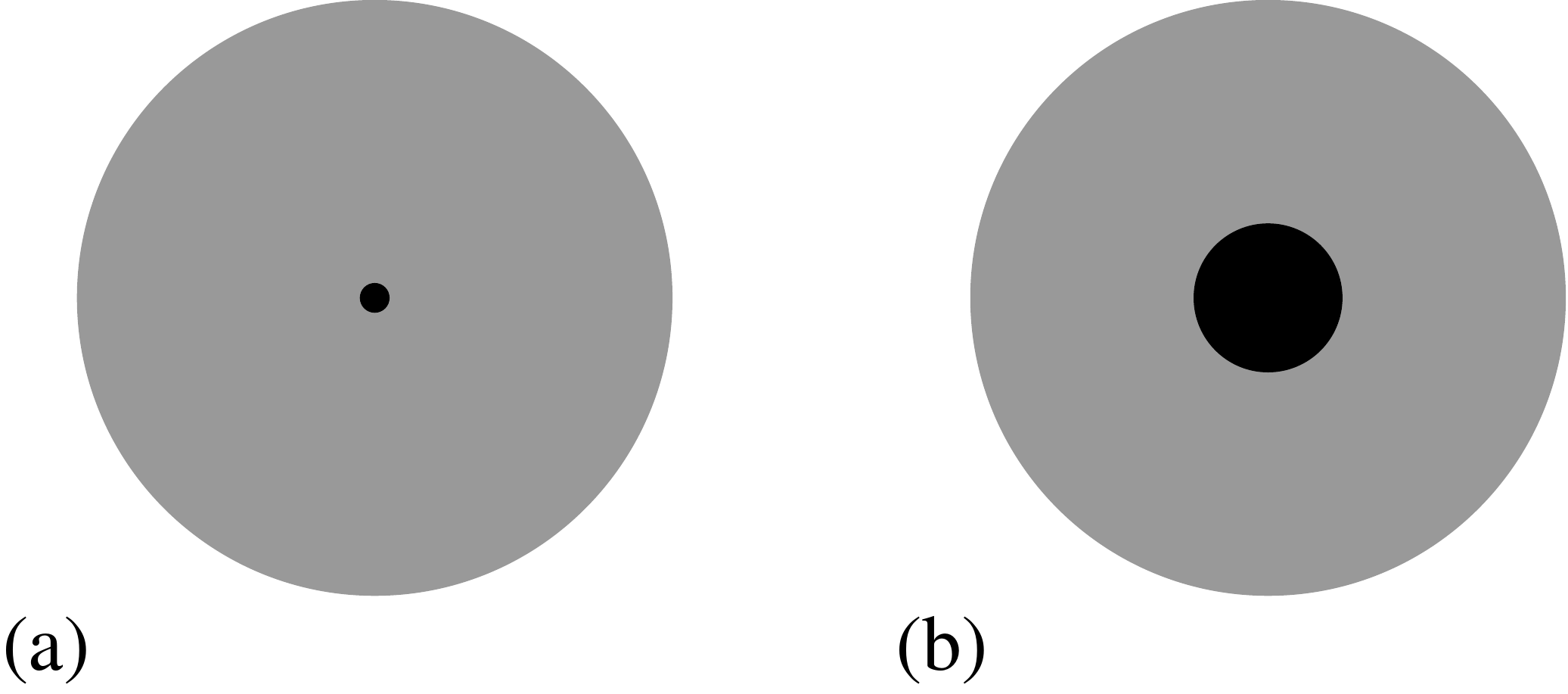}
\caption{
The static point singularity in the incompressible density
of the polar-fluid droplet 
shown in (a) as a small black disc 
induces a puncture of radius
(\ref{eq: def puncture size})
in the coordinate of the polar fluid droplet
shown in (b) as a large disc.
The polar-fluid droplet is shown as the grey disk in both
(a) and (b).
        }
\label{fig: puncture}
\end{figure}

\subsection{Winding a quasi-particle around a quasi-vortex}
\label{subsec: Winding a quasi-particle around a quasi-vortex}

We call $ \tilde{j}^{\mu} $ and $ \tilde{J}^{\mu\nu} $ the
quasi-particle and quasi-vortex currents, respectively.  We are first
going to show how they can be related to a point-like defect such as
the one represented by Eq.~(\ref{eq: monopole charge qe}), to which
the charge $e^{*}$ is associated, or the string-like defect
such as the one represented by (\ref{eq: magnetic flux phim}), to
which the charge $s^{*}$ is associated. We will then derive
the Berry phase induced when a
quasi-particle excitation winds adiabatically $n$ times around a
quasi-vortex excitation of the incompressible polar fluid with the BF
action~(\ref{eq: def BF L}). In doing so, we are going to derive the
quantization condition
\begin{equation}
\frac{g}{\hbar}\,
\frac{e^*}{e^{\ }}\,
\frac{s^*}{s^{\ }}\, n
\in\mathbb{Z}.
\end{equation}

To this end, we define
the action of the fields $b$ and $d$
interacting with the quasi-particle and quasi-vortex currents by
\begin{subequations}
\label{eq: def quantum BF + quasi-particles}
\begin{align}
&
S[b,d,\tilde{j},\tilde{J}]:= 
\int\mathrm{d}^{4}y\, 
\left(
\mathcal{L}^{\,}_{\mathrm{BF}}
+
\mathcal{L}^{\,}_{e^{*}}[\tilde{j}]
+
\mathcal{L}^{\,}_{s^{*}}[\tilde{J}]
\right),
\label{eq: def quantum BF + quasi-particles a}
\\
&
\mathcal{L}^{\,}_{\mathrm{BF}}:=
\frac{g}{2\pi}\,
\epsilon^{\mu\nu\lambda\sigma}\,
b^{\,}_{\mu\nu}\,
\partial^{\,}_{\lambda}
d^{\,}_{\sigma},
\label{eq: def quantum BF + quasi-particles b}
\\
&
\mathcal{L}^{\,}_{e^{*}}[\tilde{j}^{\mu}]:=
\frac{e^{*}}{e^{\ }}\,g\,d^{\,}_{\mu}\,
\tilde{j}^{\mu},
\label{eq: def quantum BF + quasi-particles c}
\\
&
\mathcal{L}^{\,}_{s^{*}}[\tilde{J}^{\mu\nu}]:=
\frac{s^{*}}{s^{\ }}\,
g\,
b^{\,}_{\mu\nu}\,
\tilde{J}^{\mu\nu}.
\label{eq: def quantum BF + quasi-particles d}
\end{align}
The quasi-particle and quasi-vortex currents
$\tilde{j}^{\mu}$
and
$\tilde{J}^{\mu\nu}$
couple to the fields $d^{\,}_{\mu}$
and
$
b^{\,}_{\mu\nu}
$,
respectively. 
The quasi-particle current
$\tilde{j}^{\mu}$
couples to the dynamical field $d^{\,}_{\mu}$
as the dynamical conserved current $j^{\mu}$
defined in Eq.~(\ref{eq: conserved jmu current})
does to the external electromagnetic field
$A^{\,}_{\mu}$ through the electric charge $e$
in Eq.~(\ref{eq: def e and s}). Hence, the quasi-particle charge
$e^{*}$ shares the same dimension as the electric
charge $e$, 
even though we allow for the possibility that they differ in value.
Similarly, the quasi-vortex current
$\tilde{J}^{\mu\nu}$
couples to the dynamical field $b^{\,}_{\mu\nu}$
as the dynamical conserved current 
$J^{\mu\nu}$
does to the external vortex field
$B^{\,}_{\mu\nu}$ through the vortex charge $s$
in Eq.~(\ref{eq: def e and s}). Hence, the vortex charge
$s^{*}$ shares the same dimension as $s$,
even though we allow for the possibility that they differ in value.
The path integral
\begin{align}
Z[\tilde{j},\tilde{J}]:=&\,
\int\mathcal{D}[d]\,
\int\mathcal{D}[b]\,
e^{+\mathrm{i}S[b, d, \tilde{j},\tilde{J}]/\hbar}
\nonumber\\
\equiv&\,
Z[0,0]\,
e^{+\mathrm{i}S_{\textrm{eff}}[\tilde{j},\tilde{J}]/\hbar}
\label{eq: def quantum BF + quasi-particles e}
\end{align} 
\end{subequations}
defines the quantum theory with the action 
(\ref{eq: def quantum BF + quasi-particles a})
in the background of the sources
$\tilde{j}^{\mu}$ and $\tilde{J}^{\mu\nu}$.
Their mutual interactions are captured by the effective
action 
$S^{\,}_{\mathrm{eff}}[\tilde{j},\tilde{J}]$
obtained after integrating out the $b$ and $d$ fields.

Since (\ref{eq: def quantum BF + quasi-particles a}) describes a quadratic
action, we can obtain 
$
S_{\textrm{eff}}[\tilde{j},\tilde{J}]
$
by expressing the dependence of the fields $b$ and $d$
on the currents $\tilde{j}$ and $\tilde{J}$ via the equations of
motion, which read
\begin{equation}
\frac{1}{2\pi}\,
\epsilon^{\mu\nu\lambda\sigma}\,
\partial^{\,}_{\nu}b^{\,}_{\lambda\sigma}=
-
\frac{e^{*}}{e^{\ }}
\tilde{j}^{\mu}
\label{eq: eqm from dmu}
\end{equation}
(when varying with respect to 
$d^{\,}_{\mu}$ 
for $\mu=0,1,2,3$)
and
\begin{equation}
\frac{1}{2\pi}\,
\epsilon^{\mu\nu\lambda\sigma}\,
\partial^{\,}_{\lambda}\,
d^{\,}_{\sigma}=
-
\frac{s^{*}}{s^{\ }}\,
\tilde{J}^{\mu\nu}
\label{eq: eqm from bmunu}
\end{equation}
(when varying with respect to $b^{\mu\nu}$
for $\mu,\nu=0,1,2,3$). 

Replacing the equations of motion
(\ref{eq: eqm from dmu}) and (\ref{eq: eqm from bmunu})
into (\ref{eq: def quantum BF + quasi-particles a}) yields
\begin{align}
S^{\,}_{\mathrm{eff}}[\tilde{j},\tilde{J}]=&\,
\frac{e^{*}}{e^{\ }}\,
g\,
\int\,\mathrm{d}^{4}y\,
\tilde{j}^{\mu}(y)\,d_{\mu}(y)
\nonumber\\
=&\,
-
\frac{e^{*}}{e^{\ }}\,
\frac{s^{*}}{s^{\ }}\,
g\,
\iint\mathrm{d}^{4}y\,\mathrm{d}^{4}y'\,\times
\nonumber\\
&\,
\tilde{j}^{\mu}(y)\,
\Big(
\frac{1}{2\pi}\,
\varepsilon^{\alpha\beta\lambda\mu}\,\partial_{\lambda}
\Big)^{-1}(y - y')\,
\tilde{J}^{\alpha\beta}(y').
\end{align}


We define the static point defect
\begin{equation}
\tilde{j}^{0}(t,\bs{y}):=
\delta(y^{\,}_{1})\,
\delta(y^{\,}_{2})\,
\delta(y^{\,}_{3}),
\qquad
\tilde{j}^{i}(t,\bs{y}):= 0,
\label{eq: def quasi-particle as point defect}
\end{equation}
with $i=1,2,3$. According to Eq.~(\ref{eq: eqm from dmu}),
this is the source for the static field configuration
\begin{equation}
b^{\,}_{jk}(t,\bs{y})=
-
\frac{1}{4}\,
\frac{e^{*}}{e^{\ }}\,
\epsilon^{\,}_{jki}\,
\frac{y^{\,}_{i}}{|\bs{y}|^3}
\label{eq: solution eqm for point charge at origin}
\end{equation}
with $j,k=1,2,3$.
For any closed surface $\Sigma$ that is the boundary of an open
neighborhood that contains the origin $\bm{y}=0$ and is oriented
outwards, Gauss law gives
\begin{equation}
\frac{1}{2\pi}
\iint\limits_{\Sigma}
\mathrm{d}y^{\,}_{j}\,
\mathrm{d}y^{\,}_{k}\,
b^{\,}_{jk}(t,\bm{y})=
-
\frac{e^{*}}{e^{\ }}.
\end{equation} 
Hence, the static point defect%
~(\ref{eq: def quasi-particle as point defect})
binds the monopole-like field
(\ref{eq: solution eqm for point charge at origin})
with the monopole charge $-e^{*}/e$. We may then
identify $-e^{*}/e$ with $q^{\,}_{e}$ in
Eq.~(\ref{eq: monopole charge qe}).

We define the static line defect
\begin{equation}
\tilde{J}^{03}(t,\bs{y}):=
\delta(y^{\,}_{1})\,\delta(y^{\,}_{2}),
\qquad
\tilde{J}^{\mu\nu}(t,\bs{y}):=0
\label{eq: def quasi-vortex as line defect}
\end{equation}
with $\mu=1,2,3$ and $\nu=0,1,2$.
According to Eq.~(\ref{eq: eqm from bmunu}),
this is the source for the static field configuration
\begin{subequations}
\label{eq: solution eqm for string like at origin}
\begin{align}
&
d^{\,}_{1}(t,\bs{y})=
+
\frac{s^{*}}{s^{\ }}\,
\frac{
y^{\,}_{2}
     }
     {
y^{2}_{1}
+
y^{2}_{2}
     },
\\
&
d^{\,}_{2}(t,\bs{y})=
-
\frac{s^{*}}{s^{\ }}\,
\frac{
y^{\,}_{1}
     }
     {
y^{2}_{1}
+
y^{2}_{2}
     },
\\
&
d^{\,}_{3}(t,\bs{y})=0.
\end{align}
\end{subequations}
For any closed curve $C^{\,}_{3}$ that winds around the axis
$y^{\,}_{1}=y^{\,}_{2}=0$ counterclockwise,
\begin{equation}
\frac{1}{2\pi}\,
\oint\limits_{C^{\,}_{3}}\mathrm{d}y^{\,}_{i}\,d^{\,}_{i}(t,\bm{y})=
-
\frac{s^{*}}{s^{\ }}.
\end{equation}
Hence, the static line defect%
~(\ref{eq: def quasi-vortex as line defect})
binds the field 
(\ref{eq: solution eqm for string like at origin})
of an infinitesimal magnetic flux tube running
along the $y^{\,}_{3}$ axis, i.e., a vortex field,
of flux $-s^{*}/s$.
We may then identify $-s^{*}/s$ with $q^{\,}_{s}$ in
Eq.~(\ref{eq: magnetic flux phim}).

As a quasi-particle located at the time-dependent position 
$\bs{r}(t)$ and carrying the current
\begin{equation}
\tilde{j}^{\mu}_{\mathrm{adia}}(t,\bm{y}):=
\begin{pmatrix}
\delta\big(\bm{y}-\bm{r}(t)\big)
\\
\frac{\mathrm{d}\bm{r}(t)}{\mathrm{d}t}\,
\delta\big(\bm{y}-\bm{r}(t)\big)
\end{pmatrix}
\label{eq: adiabatic Ansatz for tilde j}
\end{equation}
winds $n$ times adiabatically around the static
quasi-vortex 
(\ref{eq: def quasi-vortex as line defect}),
it acquires the Berry phase defined by 
\begin{equation}
e^{\mathrm{i}\Theta^{\,}_{\mathrm{B}}(n)/\hbar}:=
e^{\mathrm{i}S^{\,}_{\mathrm{eff}}
[\tilde{j}^{\,}_{\mathrm{adia}},\tilde{J}^{\,}_{\mathrm{adia}}]/\hbar}.
\end{equation}
The computation of $\Theta^{\,}_{\mathrm{B}}$ gives
\begin{align}
\Theta^{\,}_{\mathrm{B}}(n)=&\,
+
\frac{e^{*}}{\hbar}\,
\int\,\mathrm{d}^{4}y\,
d^{\,}_{\mu}\,
\tilde{j}^{\mu}_{\mathrm{adia}}
\nonumber\\
=&\,
-
\frac{e^{*}}{\hbar}\,
\int\mathrm{d}t\,\mathrm{d}^{3}\bs{y}\,
\sum_{i=1,2}
d^{\,}_{i}(\bs{y})\,
\tilde{j}^{\,}_{i}(t,\bs{y})
\nonumber\\
=&\,
-
\frac{e^{*}}{\hbar}\,
\int\mathrm{d}t\,\mathrm{d}^{3}\bs{y}\,
\sum_{i=1,2}
d^{\,}_{i}(\bs{y})\,
\frac{
\mathrm{d}\,r^{\,}_{i}(t)
     }
     {
\mathrm{d}\,t
     }
\,
\delta(\bs{y}-\bs{r}(t))
\nonumber\\
=&\,
-
\frac{e^{*}}{\hbar}\,
\oint\limits_{C^{\,}_{3}}\,
\mathrm{d}\bs{r}
\cdot
\bs{d}(\bs{r})
\nonumber\\
=&\,
2\pi\,
\frac{g}{\hbar}\,
\frac{e^{*}}{e^{\ }}\,
\frac{s^{*}}{s^{\ }}\,n.
\end{align}
We used Eq.~(\ref{eq: solution eqm for string like at origin})
to deduce the second and last equalities.

If we demand that the quantum theory
(\ref{eq: def quantum BF + quasi-particles})
is invariant under this adiabatic process,
we must impose the quantization condition
\begin{equation}
\label{eq: quantification condition}
\frac{g}{\hbar}\,
\frac{e^{*}}{e^{\ }}\,
\frac{s^{*}}{s^{\ }}\,n
=m
\in\mathbb{Z}.
\end{equation}

We can use this quantization condition to find the minimum possible
quantized charges in the theory. Physically we should demand that the
Berry phase be an integer multiple of $2\pi$ whenever any
quasiparticle winds once ($n=1$) around a fundamental vortex (of vorticity
$s$). Similarly, the Berry phase associated with winding once a quasivortex
around a fundamental charge (of charge $e$) must also be $2\pi$. This
yields the conditions that the minimum fractional charges and
vorticities are
\begin{equation}
\frac{e^{*}_{\rm min}}{e^{\ }}=
\frac{\hbar}{g}
\qquad
{\rm and}
\qquad
\frac{s^{*}_{\rm min}}{s^{\ }}=
\frac{\hbar}{g}
\;.
\end{equation}
This result is obtained using the minimum $m=1$.

\section{Density operator algebra and the BF theory}
\label{sec: Density operator algebra from the BF theory}

In Sec.\ \ref{subsec: Winding a quasi-particle around a quasi-vortex}, 
we extracted the braiding statistics of topological excitations 
in a polar fluid. We now deduce another important property
of a polar fluid 
namely the algebra obeyed by the density operators of the polar fluid.

We recall that, in the two-dimensional quantum Hall fluid, the particle
density operator obeys the GMP algebra (also known as the $W^{\,}_{\infty}$
algebra or the Fairlie-Fletcher-Zachos algebra).%
~\cite{GMP1985,GMP1986,Fairlie1989a,Fairlie1989b} 
The GMP algebra plays an
important role in the theory of the quantum Hall fluid.  In the
fractional quantum Hall effect, the GMP algebra can be used to
construct, via a single-mode approximation, the magneto-roton
excitation, a dispersing gapped charge-neutral collective excitation
above the ground state.  In the presence of a boundary (an edge), the
GMP algebra describes the gapless edge excitations of quantum Hall liquid. 
(To be more precise, to describe edge states one needs to consider the GMP algebra
with a central extension.  The resulting algebra is called the
$W^{\,}_{1+\infty}$ algebra.)
\cite{Iso1992, Cappelli1993, Cappelli1995, Stone91, Martinez1993, Azuma94}

In the polar fluid, it is natural to discuss, in addition to the
particle density operator, a density operator associated to the
vorticity, and commutation relations between these density operators.
The BF Lagrangian (\ref{eq: def BF L}) together with the
identification of conserved densities (currents), Eqs.\ (\ref{eq:
  conserved jmu current}) and (\ref{eq: conserved Jmunu current}),
suggest a non-vanishing commutator between these densities.  (See
below.) In this section, we discuss this issue with the help of a
fermionic microscopic model -- a free massive Dirac fermion in (3+1)
dimensions.  In the following, we will identify the density operators
associated to the particle number and the vorticities within the Dirac
model.  Assuming the large mass gap, we will then project these
density operators to the occupied bands and compute the commutation
relations.  Finally, we will make a comparison with the effective BF
theory description.

\subsection{The density algebra in the massive Dirac fermion model}
\label{subsec: The density algebra in the massive Dirac fermion model}

The Dirac Hamiltonian in question is given by  
\begin{subequations}
\label{eq: def Dirac Hamiltonian}
\begin{equation}
\hat{H}:=
\sum_{\boldsymbol{k}} 
\hat{\Psi}^{\dag}(\boldsymbol{k})\,
\mathcal{H}(\boldsymbol{k})\,
\hat{\Psi}(\boldsymbol{k}), 
\label{eq: def Dirac Hamiltonian a}
\end{equation}
where $\hat{\Psi}(\boldsymbol{k})$ 
is a four component fermion annihilation operator,
\begin{equation} 
\hat{\Psi}(\boldsymbol{k}):=
\left(
\hat{\psi}^{\,}_{1}(\boldsymbol{k}),
\hat{\psi}^{\,}_{2}(\boldsymbol{k}),
\hat{\psi}^{\,}_{3}(\boldsymbol{k}), 
\hat{\psi}^{\,}_{4}(\boldsymbol{k})
\right)^{\mathsf{T}},
\label{eq: def Dirac Hamiltonian b}
\end{equation} 
the momentum $\boldsymbol{k}\in\mathbb{R}^{3}$,
the single-particle Hermitian $4\times4$ matrix $\mathcal{H}$ takes the form
\begin{equation}
\mathcal{H}(\boldsymbol{k}):=
\sum_{i=1}^{3} 
k^{\,}_{i}\,
\alpha^{\,}_{i}
+
m\,
\beta,
\label{eq: def Dirac Hamiltonian c}
\end{equation}
and the gamma matrices are chosen to be in the Dirac representation  
\begin{equation}
\alpha^{\,}_{i}\equiv\gamma^{\,}_0\,\gamma^{\,}_{i}:=
\begin{pmatrix} 0 & \sigma^{\,}_{i} \\ \sigma^{\,}_{i} & 0 \end{pmatrix},
\qquad 
\beta\equiv\gamma^{\,}_{0}:=
\begin{pmatrix} \sigma^{\,}_{0} & 0 \\ 0 &-\sigma^{\,}_{0} 
\end{pmatrix}.
\label{eq: def Dirac Hamiltonian d} 
\end{equation}
\end{subequations}

The spectrum of $\mathcal{H}$ consists of 
two doubly degenerate bands with the energy eigenvalues 
\begin{equation}
\varepsilon^{\,}_{\pm}(\boldsymbol{k})=
\pm\sqrt{\boldsymbol{k}^{2}+m^{2}}. 
\end{equation} 
In the following, we assume the chemical potential 
such that the lowest two bands are fully
occupied and the mass gap is large, 
much larger than any perturbations that we could add to the Dirac Hamiltonian.
We are after the physics encoded by the lower bands.  
In particular, we seek the algebra obeyed by the 
charge and vortex density operators projected onto the lower bands. 
The charge-density operator in the Dirac model (before projection) 
is given by 
\begin{equation}
\hat{\rho}:= 
\hat{\Psi}^{\dag}\,\hat{\Psi}. 
\end{equation}
Once projected onto the two fully filled lowest bands, 
this operator should be compared
with $\epsilon^{ijk} \partial^{\,}_{i}\hat{b}^{\,}_{jk}$ in the BF theory. 
As for the counterpart of $\epsilon^{ijk} \partial^{\,}_{j}\hat{d}^{\,}_{k}$, 
the spin-density operator is not appropriate as spin
is not conserved due to the spin-orbit coupling. Instead,
we consider the curl of the Dirac current,
\begin{equation}
\hat{\Lambda}^{i}:=
\epsilon^{ijk}\partial^{\,}_{j}
\left(
\hat{\Psi}^{\dag}
\,\gamma^{0}\,\gamma_{k}\,
\hat{\Psi}
\right). 
\end{equation}
Assuming the mass $m$ to be ``large'', 
we then evaluate the commutator for 
the charge and vortex density operators 
projected onto the lowest two occupied bands.  

The comparison between the BF field theory and the non-interacting
Dirac model is not expected to be perfect.  To elaborate this point,
we go momentarily back to two spatial dimensions.  On the one hand,
the GMP algebra is obtained for the charge-density operator projected
onto the lowest Landau level, whereby the lowest Landau level has a
uniform Berry curvature. On the other hand, the projected
charge-density operator in two-dimensional Chern insulators do not
obey the GMP algebra, since Chern bands have a non-uniform Berry
curvature related as they are to the massive Dirac Hamiltonian in
two-dimensional space.%
~\cite{Parameswaran12,Goerbig12,Estienne12,Murthy12,Murthy11,Chamon12}

While it may be possible to use three-dimensional Landau levels
to make a better comparison with the density algebra derived 
from the BF theory,
we will stick with the Dirac model for the sake of simplicity.
A ``trick'' that we will use to improve the comparison is 
that we will focus on the region of the momentum space
$|\boldsymbol{k}|/m \ll 0$ 
for which the Berry curvature is asymptotically uniform. 

The projection onto the lowest bands can be done by first transforming
the fermion operators 
$\hat{\psi}^{\,}_{\alpha}(\boldsymbol{k})$ 
with $\alpha=1,2,3,4$
into  the eigenoperators $\hat{\chi}^{\,}_{a}(\boldsymbol{k})$
with $a=1,2,3,4$
of the Hamiltonian $\mathcal{H}(\boldsymbol{k})$ according to 
\begin{equation}
\hat{\psi}^{\dag}_{\alpha}(\boldsymbol{k})=
\sum_{b=1}^{4}
u^{b*}_{\alpha}(\boldsymbol{k})\,
\hat{\chi}^{\dag}_{b}(\boldsymbol{k}),
\end{equation}
where $u^{b}_{\alpha}(\boldsymbol{k})$ are the components of the
eigenfunctions (Bloch wave function) of $\mathcal{H}(\boldsymbol{k})$.
In terms of $\hat{\chi}$ and $u$, the projected charge-density operator 
with momentum $\boldsymbol{q}$ is 
\begin{equation}
\tilde{\rho}(\boldsymbol{q}):=
\sum_{\boldsymbol{k}}
\sum_{\alpha=1}^{4}
\sum_{a,b=1}^{2}
\left[
u^{*}_{\alpha}(\boldsymbol{k})\, 
u^{\ }_{\alpha}(\boldsymbol{k}+\boldsymbol{q})
\right]^{ab}
\hat{\chi}^{\dag}_{a}(\boldsymbol{k})\,
\hat{\chi}^{\ }_{b}(\boldsymbol{k}+\boldsymbol{q}).
\end{equation}
Projected operators acquire the $\tilde{}$ symbol instead of the 
$\hat{}$ symbol in order to imply the summation convention
$\alpha=1,2,3,4$ on the Dirac labels, whereas the summation convention
is restricted to the labels for the occupied Bloch bands, i.e.,
$a,b=1,2$.
For $\boldsymbol{q}\to\boldsymbol{0}$, 
we expand $u(\boldsymbol{k}+\boldsymbol{q})$ 
to linear order in $\boldsymbol{q}$. Summing over the Dirac
indices $\alpha=1,\cdots,4$ gives
\begin{subequations} 
\begin{equation}
\tilde{\rho}(\boldsymbol{q})\approx
\sum_{\boldsymbol{k}} 
\sum_{a,b=1}^{2}
\left[
1+q^{i}\,A^{\,}_{i}(\boldsymbol{k})
\right]^{ab}
\hat{\chi}^{\dag}_{a}(\boldsymbol{k})\,
\hat{\chi}^{\ }_{b}(\boldsymbol{k}+\boldsymbol{q}), 
\end{equation}
where 
\begin{equation} 
A^{i}(\boldsymbol{k}):=
\sum_{\alpha=1}^{4}
u^{* }_{\alpha}(\boldsymbol{k})\,
(\partial^{i}u^{\,}_{\alpha})(\boldsymbol{k}),
\qquad
i=1,2,3,
\end{equation} 
is a non-Abelian U(2) Berry connection
and the summation convention over the index
$i=1,2,3$ that labels the components of the
three-dimensional wave number $\bm{q}$ is implied.
This non-Abelian U(2) Berry connection
can be decomposed into a U(1) part ($\bm{A}^{\,}_{1}$) 
and an SU(2) part ($\bm{A}^{\,}_{2}$).
For the massive Dirac Hamiltonian in $(3+1)$-dimensional
space and time, their components labeled by $i=1,2,3$ are
\begin{equation}
(A^{i}_{1})^{ab}(\boldsymbol{k})=
\frac{-k^{i}}{2k^{\,}_{0}\,(k^{\,}_{0}+m)}\,\delta^{ab},
\end{equation}
and
\begin{equation}
(A^{i}_{2})^{ab}(\boldsymbol{k})=
\frac{
\mathrm{i}\epsilon^{\,}_{ijk}\,(\sigma^{j})^{ab}\,k^{k}
     }
     {
2k^{\,}_{0}\,(k^{\,}_{0}+m)
     }, 
\end{equation}
\end{subequations} 
respectively, where $k^{\,}_{0}:=\sqrt{\boldsymbol{k}^{2}+m^{2}}$.

Similarly, the components labeled by the index $i=1,2,3$ of the
spin-density operator are defined to be
\begin{subequations}
\begin{equation}
\hat{\Lambda}^{i}(\boldsymbol{q}):=
\epsilon^{ijk}\,\partial^{j}j^{k}(\boldsymbol{q}),
\end{equation}
where $\hat{j}^{i}(\boldsymbol{q})$ is the Dirac 3-current operator
\begin{equation}
\hat{j}^{i}(\boldsymbol{q}):=
\sum_{\boldsymbol{k}} 
\hat{\bar{\Psi}}(\boldsymbol{k})
\gamma^{i}
\hat{\Psi}(\boldsymbol{k}+\boldsymbol{q}). 
\end{equation}
\end{subequations}
After projecting onto the lowest two occupied bands, 
the spin-density operator takes the form
\begin{align}
\tilde{\Lambda}^{i}(\boldsymbol{q}):=&\,
\sum_{\boldsymbol{k}} 
\mathrm{i}
\epsilon^{ijk}\,
q^{\,}_{j}
\left[
u^{*}_{\alpha}
\left(\gamma^{0}\gamma^{k}\right)^{\alpha\beta}
u^{\,}_{\beta}(\boldsymbol{k}+\boldsymbol{q})
\right]^{ab}
\nonumber\\
&\,
\times 
\hat{\chi}^{\dag}_{a}(\boldsymbol{k})\,
\hat{\chi}^{\,}_{b}(\boldsymbol{k}+\boldsymbol{q}).
\end{align}
To lowest leading order in a gradiant expansion of the 
Bloch states $a,b=1,2$,
\begin{subequations}
\begin{align}
\tilde{\Lambda}^{i}(\boldsymbol{q})\approx&\,
\sum_{\boldsymbol{k}} 
\mathrm{i}
\epsilon^{ijk}\,
q^{\,}_{j}\,
\left[
B^{\,}_{0,k}(\boldsymbol{k})
+
q^{l}\,
(B_{1,k}^{l}(\boldsymbol{k})
+
B_{2,k}^{l}(\boldsymbol{k}))
\right]^{ab}
\nonumber\\
&\,
\times 
\hat{\chi}^{\dag}_{a}(\boldsymbol{k})\,
\hat{\chi}_{b}(\boldsymbol{k}+\boldsymbol{q}). 
\end{align}
For the massive Dirac Hamiltonian in $(3+1)$ dimensional space and time, 
\begin{align}
&
B^{\,}_{0,i}(\boldsymbol{k})=
\frac{k^{\,}_{i}}{k^{\,}_{0}}\,
\delta^{ab},
\\
&
B^{i}_{1,j}(\boldsymbol{k})=
\frac{-k^{i}\,k_{j}\,m}{2\,k^{3}_{0}\,(k^{\,}_{0}+m)}\,
\delta^{ab},
\\
&
B^{i}_{2,j}(\boldsymbol{k})=
\mathrm{i}
\left[
\epsilon^{\,}_{ljl}\,
\frac{k^{i}\,k^{l}}{k^{\,}_{0}}
+
\epsilon^{\,}_{jil}
(k^{\,}_{0}+m) 
\right]\,
(\sigma^{l})^{ab}.
\end{align}
\end{subequations}
Again, we have explicitly kept terms that vanish by contraction with an 
antisymmetric tensor.

If we only consider the leading order term in an expansion in powers 
of the components of $q^{\,}_{1}$ and $q^{\,}_{2}$, we obtain
\begin{align}
\left[
\tilde{\rho}(\boldsymbol{q}^{\,}_{1}),
\tilde{j}^{\,}_{i}(\boldsymbol{q}^{\,}_{2})
\right]=&\,
q^{j}_{1}
\sum_{\boldsymbol{k}} 
\left[\partial^{\,}_{j}B^{\,}_{0i}\right]^{ab}
\hat{\chi}^{\dag}_{a}(\boldsymbol{k})\,
\hat{\chi}^{\,}_{b}
(\boldsymbol{k}+\boldsymbol{q}^{\,}_{1}+\boldsymbol{q}^{\,}_{2})
\nonumber\\
&\,
+
\cdots
\end{align}
and
\begin{align}
\left[
\tilde{\rho}(\boldsymbol{q}^{\,}_{1}),
\tilde{\Lambda}^{\,}_{i}(\boldsymbol{q}^{\,}_{2})
\right]=&\,
\mathrm{i}\epsilon^{\,}_{ijk}\,
q^{l}_{1}\,
q^{j}_{2}
\sum_{\boldsymbol{k}} 
\left[
\partial^{\,}_{l}B^{k}_{0}(\boldsymbol{k})
\right]^{ab}
\nonumber\\
&\,\times
\hat{\chi}^{\dag}_{a}(\boldsymbol{k})\,
\hat{\chi}^{\,}_{b}(\boldsymbol{k}+\boldsymbol{q}^{\,}_{1}+\boldsymbol{q}^{\,}_{2})
+ 
\cdots. 
\end{align}
When $|\boldsymbol{k}|\ll m$, we arrive at 
\begin{align}
\left[
\tilde{\rho}(\boldsymbol{q}^{\,}_{1}),
\tilde{\Lambda}^{\,}_{i}(\boldsymbol{q}^{\,}_{2})
\right]&= 
\mathrm{i}\epsilon^{\,}_{ijk}\,
\frac{q^{k}_{1}\,q^{j}_{2}}{m}
\tilde{\rho}(\boldsymbol{q}^{\,}_{1}+\boldsymbol{q}^{\,}_{2})
+
\cdots.  
\label{bf algebra, dirac}
\end{align}
This is an analogue of the GMP algebra.

We now compare the
commutator (\ref{bf algebra, dirac})
derived from the massive Dirac model 
with the corresponding 
commutator in the BF theory. 
We begin with the BF Lagrangian density
(\ref{eq: def BF L})
in the temporal gauge
\begin{subequations} 
\begin{equation}
d^{\,}_{0}=b^{\,}_{0i}=0.
\end{equation}
It is given by
\begin{equation}
\mathcal{L}= 
\frac{g}{2\pi}\,
\epsilon^{\,}_{ijk}\,
\dot{d}^{\,}_{i}\,
b^{\,}_{jk}=  
g\,\dot{d}^{\,}_{i}\,B^{i}, 
\end{equation}
where we have defined 
\begin{equation}
B^{i}:=
\frac{
1
     }
     {
2\pi
     }\,
\epsilon^{ijk}\,
b^{\,}_{jk}
\equiv
\frac{
1
     }
     {
2\pi
     }\,
\epsilon^{\,}_{ijk}\,
b^{\,}_{jk}.
\end{equation}
\end{subequations} 
Canonical quantization 
for the canonical pair $d^{\,}_{i}$ and $g\,B^{i}$
implies the equal-time commutation relation
\begin{equation}
\left[
\hat{d}^{\,}_{i}(\boldsymbol{x}),
\hat{B}^{j}(\boldsymbol{y})
\right]=
\mathrm{i} 
g^{-1}\,
\delta(\boldsymbol{x}-\boldsymbol{y})\,
\delta^{j}_{\ i} 
\label{BF commutator}
\end{equation}
for $i,j=1,\cdots,3$.
Recalling the definitions of the conserved currents,  
Eqs.\ (\ref{eq: conserved jmu current}) and (\ref{eq: conserved Jmunu current}), 
the commutator (\ref{BF commutator}) resembles 
the commutator (\ref{bf algebra, dirac}),
i.e., the presence of the factor $\epsilon^{\,}_{ijk}\,q^{k}_{1}\,q^{j}_{2}$,
although there is no
particle number density operator on the right-hand side
of the commutator (\ref{BF commutator}). 

In fact, the absence of the density operator on the right-hand side of 
of Eq.\ (\ref{BF commutator}) is anticipated (see below),  
and the comparison between the commutators derived from the microscopic model
and from the effective field theory is not expected to be complete. 
Within the BF theory description,  
the particle density is completely frozen in the bulk and does not fluctuate.
Hence, the density operator on the right-hand side of the commutator 
(\ref{BF commutator}) 
is ``invisible''. 
This situation is completely analogous to the Chern-Simons description 
of the quantum Hall fluid.
In the Chern-Simons description of quantum Hall fluid, the only
collective charge fluctuations described by the Chern-Simons theory
are edge excitations (apart from the point-like quasiparticle
excitations in the bulk). Hence, one can not derive the GMP
algebra in the bulk from the Chern-Simons theory.  Nevertheless, the
description of edge excitations derived from the Chern-Simons theory
is consistent with the edge excitations derived from the GMP algebra. 
\cite{Iso1992, Cappelli1993, Cappelli1995, Stone91, Martinez1993, Azuma94}

\section{Discussion}
\label{sec:Discussion}

We have formulated a hydrodynamic description of gapped topological
electron fluid in term of the BF effective field theory.  Just as
fluid dynamics is an efficient description of a collection of
macroscopic number of interacting particles, the hydrodynamic BF field
theory allows us to describe incompressible electron liquid beyond
single particle physics.  From the BF theory, we have extracted
statistical information of defects in the polar fluid.

We close with two comments.
(i)
In the last section,
we have linked the hydrodynamic BF theory 
to the algebra of densities in the polar fluid. 
The hydrodynamic BF theory may be derived, alternatively,
by using functional bosonization techniques. 
In the functional bosonization approach, 
one derives an effective action 
that encodes the low-energy and long-wavelength properties of 
conserved quantities (hydrodynamic modes) for a given microscopic model.  
For example, descriptions of topological insulators in terms of
effective field theories have been derived by bosonizing 
the charge $U(1)$ degrees of freedom in topological insulators.%
~\cite{Chan12}
In the polar fluid, we are concerned with two kinds of densities, 
the charge and vorticity densities. 
A functional bosonization can be adopted to take into 
account both kinds of densities.%
~\cite{unpublished}

(ii) The purpose of the present paper was to derive a hydrodynamic
description of incompressible topological fluid with a few basic
assumptions.  As such, hydrodynamic field theories can describe both
bosonic and fermionic lattice models, e.g., bosonic and fermionic
topological insulators, at low energies and long wavelengths.  (See,
e.g., Refs.\ \onlinecite{Vishwanath2013,Kapustin2014}
and  \onlinecite{YeGu2014} for
discussions of bosonic topological insulators and their descriptions in
terms of BF theories).  By construction, hydrodynamic field theories
are written in terms of bosonic degrees of freedom (describing
conserved hydrodynamic modes). Hence, the distinction between the cases
when the underlying particles obey bosonic or fermionic statistics
has to be encoded in a rather subtle way.  For example, in the Chern-Simons
theory of the fractional quantum fluid, the distinction between
bosonic/fermionic statisctic of fundamental particles manifests itself
as eveness/oddness of the level of the Chern-Simons term.  In our
description of three-dimensional topological incompressible fluid, we
expect that the bosonic/fermionic statistics is encoded by the periodicity
of the $\theta$ angle in the axion term; for bosonic (fermionic)
underlying particles, the periodicity is $4\pi$ ($2\pi$).  

\acknowledgements 

We thank Tom Faulkner for useful discussion. 
We acknowledge
the visitor program at Perimeter Institute, 
the PCTS program ``Symmetry in Topological Phases''
at Princeton Center for Theoretical Science 
(17-18 March 2014)
and 
the international workshop  
``Topology and Entanglement in Correlated Quantum Systems''
at Max Planck Institute for the Physics of Complex Systems
(14-25 July 2014), 
where the part of this work was carried out.  
This work was partially supported by the National Science Foundation through grant DMR-1064319 (XC).
SR acknowledges support of the Alfred P. Sloan Research Foundation. 
Research at the Perimeter Institute is supported by the Government
of Canada through Industry Canada and by the Province of Ontario
through the Ministry of Economic Development and Innovation. (L.H.S.)

\end{document}